\PassOptionsToPackage{colorlinks=true,citecolor=blue,linkcolor=red,urlcolor=blue}{hyperref}
\documentclass[aps,prd,floatfix,onecolumn,10pt,amsmath,amssymb,showpacs,showkeys]{revtex4-2}
\usepackage{graphicx}
\usepackage{xcolor}
\usepackage{booktabs}
\usepackage{bm}
\usepackage{hyperref}

\begin{document}
\title{Reliable Hybrid Neural Surrogates for Multidimensional Grids: Application to Double Parton Distributions}

	\preprint{}

\author{R.~Kord Valeshabadi$^{1}$}

\email[Corresponding author: ]{ramin.kord@ipm.ir}

\author{S.~Rezaie$^{1}$}
\email{somayeh@ipm.ir}

\affiliation{
	$^{1}$School of Particles and Accelerators, Institute for Research in Fundamental Sciences (IPM), P.O. Box 19568-36681, Tehran, Iran
}

\begin{abstract}
The conventional grid-based approach used for storing and evaluating ordinary parton distribution functions (PDFs), as implemented for example in LHAPDF, becomes much more demanding for double parton distributions (DPDs). While a collinear PDF depends on one longitudinal momentum fraction and one factorization scale, an unequal-scale DPD depends on two momentum fractions and two independent factorization scales for every parton-flavor combination. Direct tabulation of this four-dimensional dependence therefore requires considerably more disk space and runtime memory.

In this work, we develop a hybrid neural-network method for compact storage and fast evaluation of $y$-independent unequal-scale DPDs. The network compresses the reference grid by reproducing most of its values, while the original values are stored only at points where the neural prediction does not reach the required accuracy. In this way, the neural model captures the bulk of the grid, with a sparse table covering the remaining difficult points.

For the GS09-based dense grid studied here, only $0.694\%$ of the active flavor values need to be stored in the sparse table. Compared with the compressed dense grid, the hybrid method reduces the disk footprint by a factor of $10.6$, peak memory use by a factor of $31.6$, and initialization time by about a factor of $10.6$. At the same time, its evaluation throughput reaches about $85\%$ of the dense-grid rate. The method is implemented natively in C++ within PDFxTMDLib and requires neither Python nor PyTorch at runtime. The hybrid scheme therefore offers much lower memory and startup costs, with only a modest reduction in evaluation speed.

\end{abstract}
\maketitle

\section{Introduction}
Parton distribution functions (PDFs) are essential ingredients in theoretical predictions for hard processes involving hadrons. Within the collinear factorization framework, a hadronic cross section, such as that for proton--proton collisions at the LHC, can be expressed as a convolution of perturbatively calculable partonic cross sections with the PDFs of the incoming hadrons:
\begin{equation}
	\sigma_{pp\to X}
	=
	\sum_{i,j}
	\int_0^1 dx_1
		\int_0^1 dx_2\,
		f_i(x_1,\mu_F^2)\,
		f_j(x_2,\mu_F^2)\,
		\hat{\sigma}_{ij\to X}
	\left(
	x_1,x_2;
	\mu_F^2,\mu_R^2
	\right),
	\label{eq:sps_fact}
\end{equation}
where $f_i(x_1,\mu_F^2)$ and $f_j(x_2,\mu_F^2)$ denote the PDFs for finding partons of flavors $i$ and $j$ carrying longitudinal momentum fractions $x_1$ and $x_2$ in the two incoming protons, respectively. The quantity $\hat{\sigma}_{ij\to X}$ denotes the perturbatively calculable partonic cross section, while $\mu_F$ and $\mu_R$ are the factorization and renormalization scales.

Equation~\eqref{eq:sps_fact} describes single parton scattering (SPS), in which one parton from each incoming proton participates in a single hard interaction. In a proton--proton collision, however, two distinct hard partonic interactions may occur within the same hadronic collision. This mechanism is known as double parton scattering (DPS). In a collinear DPS description that keeps information about the relative transverse separation of the two active partons, the cross section for producing final states $A$ and $B$ can be written schematically as
\begin{equation}
	\label{eq:dps_fact}
	\begin{aligned}
		\sigma_{pp\to A+B}^{\mathrm{DPS}}
		={}&
		\frac{1}{C}
		\sum_{\substack{a_1,a_2\ b_1,b_2}}
		\int dx_1,dx_2
		\int d\bar{x}_1,d\bar{x}_2
		\int d^2\boldsymbol{y} F_{a_1a_2}
		\left(
		x_1,x_2,\boldsymbol{y};
		\mu_1^2,\mu_2^2
		\right)
		F_{b_1b_2}
		\left(
		\bar{x}_1,\bar{x}_2,\boldsymbol{y};
		\mu_1^2,\mu_2^2
		\right)
		\\
		&\times
		\hat{\sigma}_{a_1b_1\to A}
		\left(
		x_1,\bar{x}_1;
		\mu_1^2
		\right)
		\hat{\sigma}_{a_2b_2\to B}
		\left(
		x_2,\bar{x}_2;
		\mu_2^2
		\right).
	\end{aligned}
\end{equation}
Here, $F_{a_1a_2} \left(x_1,x_2,\boldsymbol{y}; \mu_1^2,\mu_2^2 \right)$ denotes a double parton distribution. It describes the correlated distribution of two partons of flavors $a_1$ and $a_2$ carrying longitudinal momentum fractions $x_1$ and $x_2$, with relative transverse separation $\boldsymbol{y}$, when probed at the factorization scales $\mu_1$ and $\mu_2$. The barred momentum fractions $\bar{x}_1$ and $\bar{x}_2$ refer to the corresponding partons in the second proton. The quantities $\hat{\sigma}_{a_1b_1\to A}$ and $\hat{\sigma}_{a_2b_2\to B}$ denote the perturbatively calculable cross sections for the two hard subprocesses.

The factor $C$ is a combinatorial symmetry factor,
\begin{equation}
	C =
	\begin{cases}
		2, & A=B,\\
		1, & A\neq B,
	\end{cases}
\end{equation}
which prevents double counting when the two hard final states are identical.

In the following, we restrict the discussion to unpolarized DPDs. In this case, the DPD is a scalar under rotations in the transverse plane and its dependence on the relative transverse separation $\boldsymbol{y}$ can be expressed in terms of its magnitude, $y=|\boldsymbol{y}|$~\cite{Diehl:2011yj,Diehl:2023cth}. We therefore write the distribution as $F_{a_1a_2}(x_1,x_2,y;\mu_1^2,\mu_2^2)$.

A commonly used approximation is to separate the transverse dependence from the longitudinal momentum and scale dependence,
\begin{equation}
	F_{a_1a_2}
	\left(
	x_1,x_2,\boldsymbol{y};
	\mu_1^2,\mu_2^2
	\right)
	\approx
	D_{a_1a_2}
	\left(
	x_1,x_2;
	\mu_1^2,\mu_2^2
	\right)
	T_{a_1a_2}(y),
	\label{eq:dpd_transverse_factorization}
\end{equation}
where $D_{a_1a_2}$ denotes the corresponding collinear DPD and $T_{a_1a_2}(y)$ describes the transverse profile of the two-parton system. This separation is itself an approximation, since perturbative parton splitting can generate correlations between the longitudinal variables and the transverse separation~\cite{Diehl:2017kgu,Diehl:2011yj}.

A further approximation, often used in phenomenological applications, is to express the collinear DPD in terms of ordinary single PDFs,
\begin{equation}
	D_{a_1a_2}^{\mathrm{fac}}
	\left(
	x_1,x_2;
	\mu_1^2,\mu_2^2
	\right)
	=
	f_{a_1}(x_1,\mu_1^2)
	f_{a_2}(x_2,\mu_2^2)
	\Theta(1-x_1-x_2),
	\label{eq:factorized_dpd}
\end{equation}
where the step function enforces the kinematic constraint $x_1+x_2\leq1$. This product ansatz is a phenomenological approximation rather than a general property of DPDs. The two partons can exhibit longitudinal momentum and flavor correlations that are not captured by a simple product of two single PDFs. Such a product also does not, in general, satisfy the DPD momentum and valence-number sum rules~\cite{Gaunt:2009re}. Even so, the factorized ansatz is a useful reference for studying genuine two-parton correlations.

The factorization scale dependence of ordinary PDFs is governed by the DGLAP evolution equations~\cite{DGLAP1,DGLAP2,DGLAP3}, whereas DPDs obey the double-DGLAP (dDGLAP) evolution equations~\cite{Ceccopieri:2010kg,Gaunt:2009re}. The dDGLAP evolution generates additional longitudinal correlations between the two partons, so independently evolved single PDFs do not, in general, reproduce the full DPD evolution.

Numerically solving these coupled integro-differential equations whenever a distribution is requested would be inefficient for applications requiring a large number of repeated evaluations. For ordinary PDFs, this problem is conventionally avoided by performing the QCD evolution beforehand and distributing the resulting distributions as numerical grids. Values at arbitrary phase-space points are then obtained by interpolation, as implemented, for example, in LHAPDF~\cite{LHAPDF6}.

For a conventional collinear PDF, the distribution depends on one longitudinal momentum fraction and one factorization scale, $f_a(x,\mu^2)$. As an illustrative example, the \texttt{MSTW2008LO} set~\cite{MSTW}, which is also used as the single-PDF input for the GS09 distributions considered in this work, is tabulated on a grid with $N_x=64$ momentum-fraction nodes and $N_\mu=45$ scale nodes for 11 parton species. The corresponding ASCII grid file occupies approximately $510.4\;\mathrm{kB}$.

The storage problem becomes much more demanding for DPDs even without transverse-distance dependence. Here we consider the $y$-independent collinear DPD, $D_{a_1a_2}(x_1,x_2;\mu_1^2,\mu_2^2)$, which depends on two longitudinal momentum fractions and two independent factorization scales. The reference grid uses $N_x=56$ nodes along each momentum-fraction axis and $N_\mu=41$ nodes along each scale axis. For $n_f=5$, the full flavor structure comprises $121$ parton-flavor combinations. The resulting reference table occupies approximately $2.6\;\mathrm{GB}$ in its ASCII representation~\cite{KordValeshabadi:2026unequal}, whereas a single-PDF set typically requires only kilobyte- to megabyte-scale storage. Keeping the reference table in memory for repeated evaluation is also costly, motivating a more compact representation even for the $y$-independent case considered here.

This increase comes from the dimensionality of the distribution. A single PDF is tabulated over $(x,\mu^2)$, whereas a general unequal-scale collinear DPD depends on $(x_1,x_2,\mu_1^2,\mu_2^2)$ for every parton-flavor combination. Direct tabulation becomes more expensive as the resolution of any axis is increased. A compact alternative must avoid storing the complete DPD table without losing the numerical accuracy of the reference calculation.

An alternative strategy for handling the high dimensionality of DPDs is implemented in the ChiliPDF library~\cite{Diehl:2023cth}. ChiliPDF describes the momentum-fraction and transverse-separation dependence with Chebyshev grids and avoids an explicit grid in the two factorization scales. DPDs at different scales are instead obtained by solving the evolution equations on the fly, which considerably reduces the memory required for a fully scale-dependent representation. However, this approach introduces an additional computational cost whenever DPDs at new scale combinations are requested. For $n_f=5$, corresponding to 121 parton-flavor combinations, the evolution of a complete DPD grid with $p_x=54$ nodes in each momentum-fraction direction and $p_y=48$ transverse-distance nodes from $\mu_1=\mu_2=15\;\mathrm{GeV}$ to $150\;\mathrm{GeV}$ requires approximately $1$--$2\;\mathrm{s}$~\cite{Diehl:2023cth}. While this timing refers to the evolution of the complete grid rather than a single DPD evaluation, repeating such calculations for many scale pairs can become restrictive in Monte Carlo applications.

These considerations motivate the search for compact representations that preserve the numerical information of multidimensional DPD grids while reducing the cost of repeated evaluations. Neural networks provide flexible nonlinear function approximators and have been used both for the parametrization of parton distributions and as surrogate models for high-dimensional numerical problems~\cite{HORNIK1989359,Ball2009NNPDF,Tripathy2018,Giacomini2026}. In this work, the network is not used to solve the DPD evolution equations or to determine distributions from experimental data. Instead, it provides a compact approximation to an already computed multidimensional DPD grid.

We construct a hybrid neural-network surrogate for compact storage and efficient evaluation of $y$-independent DPDs. The network captures the overall behavior of the reference distributions across the four-dimensional kinematic domain and all flavor channels, while predictions that do not satisfy a prescribed accuracy criterion at the original grid nodes are replaced by the corresponding reference values stored in a sparse correction table. Most of the grid is represented by the neural model, and only a small fraction of values must be stored separately. This reduces storage and memory requirements and avoids repeated dDGLAP evolution during inference while maintaining an evaluation rate comparable to that of the dense grid. Finally, it should be noted that although the present study is restricted to $y$-independent DPDs, the same strategy could in principle be extended to distributions with explicit transverse-separation dependence.

The structure of the paper is as follows. Section~\ref{sec:DPD_Models} describes the reference DPDs and numerical grids. The construction of the hybrid surrogate and its implementation in PDFxTMDLib~\cite{Valeshabadi2026} are presented in Sec.~\ref{sec:HybridNN}. Section~\ref{sec:NumVal_Perf} examines its accuracy, resource requirements, and computational performance, and Sec.~\ref{sec:Conclusions} summarizes the main results.

\section{Reference \texorpdfstring{$y$}{y}-Independent Unequal-Scale DPDs}
\label{sec:DPD_Models}

This section specifies the reference DPDs used to train and validate the surrogate developed in this work. We first summarize the equal-scale GS09 distributions and the dDGLAP evolution on which they are based. We then describe the unequal-scale construction implemented with ChromaPDFEvolver and the numerical grids used as the reference data throughout this study.

\subsection{Equal-Scale GS09 Input}
\label{subsec:GS09_equal}

The starting point of our reference construction is the GS09 DPD set~\cite{Gaunt:2009re}. GS09 provides leading-order, non-factorized collinear DPDs based on the MSTW2008LO single PDFs~\cite{MSTW} at the initial scale $Q_0^2=1\;\mathrm{GeV}^2$. The equal-scale distributions are obtained by solving the leading-order dDGLAP evolution equations. For a common factorization scale $\mu^2$, the evolution equation can be written as~\cite{Ceccopieri:2010kg,Gaunt:2009re}
\begin{align}
	\mu^2
	\frac{\partial D_{ij}(x_1,x_2;\mu^2)}
	{\partial \mu^2}
	={}&
	\frac{\alpha_s(\mu^2)}{2\pi}
	\Bigg\{
	\nonumber\\
	&\sum_k
	\int_{x_1}^{1-x_2}
	\frac{dz_1}{z_1}\,
	D_{kj}(z_1,x_2;\mu^2)\,
	P_{k\to i}
	\left(\frac{x_1}{z_1}\right)
	\nonumber\\
	&+
	\sum_l
	\int_{x_2}^{1-x_1}
	\frac{dz_2}{z_2}\,
	D_{il}(x_1,z_2;\mu^2)\,
	P_{l\to j}
	\left(\frac{x_2}{z_2}\right)
	\nonumber\\
	&+
	\sum_k
	\frac{f_k(x_1+x_2;\mu^2)}
	{x_1+x_2}\,
	P^{R}_{k\to ij}
	\left(
	\frac{x_1}{x_1+x_2}
	\right)
	\Bigg\}.
	\label{eq:equal_scale_dDGLAP}
\end{align}

The first two terms describe the homogeneous evolution of the two parton legs, whereas the last term is the non-homogeneous contribution associated with perturbative splitting of a single parent parton into the two observed partons. The GS09 DPDs incorporate nontrivial longitudinal correlations and are constructed using the DPD momentum and number sum rules as constraints~\cite{Gaunt:2009re}. These equal-scale distributions are used as the initial conditions for the unequal-scale evolution described in Sec.~\ref{subsec:unequal_scale_grids}.

\subsection{Unequal-Scale Evolution and Reference Grids}
\label{subsec:unequal_scale_grids}
The equal-scale DPDs, $D_{ij}(x_1,x_2;\mu^2)$, are obtained following the GS09 construction of Ref.~\cite{Gaunt:2009re}. To construct the corresponding unequal-scale distributions, $D_{ij}(x_1,x_2;\mu_1^2,\mu_2^2)$, we employ the unequal-scale evolution formalism of Refs.~\cite{Ceccopieri:2010kg,Gaunt:2009re}, implemented in the ChromaPDFEvolver code developed and validated in our previous work~\cite{KordValeshabadi:2026unequal}. In this approach, the equal-scale distribution at the lower of the two scales is used as the initial condition for the subsequent evolution of the parton associated with the higher scale, while the scale of the other parton remains fixed.

For example, when $\mu_2^2>\mu_1^2$, the equal-scale distribution $D_{ij}(x_1,x_2;\mu_1^2,\mu_1^2)$ provides the initial condition for the evolution of the second parton,
\begin{equation}
	\mu_2^2
	\frac{\partial D_{ij}(x_1,x_2;\mu_1^2,\mu_2^2)}{\partial \mu_2^2}
	=
	\frac{\alpha_s(\mu_2^2)}{2\pi}
	\sum_l
	\int_{x_2}^{1-x_1}\frac{dz_2}{z_2}\,
	D_{il}(x_1,z_2;\mu_1^2,\mu_2^2)\,
	P_{l\to j}\!\left(\frac{x_2}{z_2}\right).
	\label{eq:unequal_scale_dDGLAP}
\end{equation}
For $\mu_1^2>\mu_2^2$, the corresponding equation is obtained by interchanging the two parton legs. The non-homogeneous splitting contribution does not appear in this second evolution step because the splitting contribution accumulated up to the lower scale is already included in the equal-scale DPD used as the initial condition~\cite{Ceccopieri:2010kg,Gaunt:2009re,KordValeshabadi:2026unequal}.

The evolution is performed using the variable $t=\ln(\mu^2)$ and the adaptive Runge--Kutta--Fehlberg algorithm implemented in GSL~\cite{GSLRefManual3rd}. The numerical solution employs an internal longitudinal grid with $N_x^{\mathrm{solver}}=90$ points. The evolved distributions are subsequently interpolated and stored on a reduced grid with $N_x=56$ nodes along each momentum-fraction axis and $N_\mu=41$ nodes along each scale axis, covering $10^{-6}\leq x<1$ and $1\;\mathrm{GeV}^2\leq\mu^2\leq10^9\;\mathrm{GeV}^2$.

These tables are stored in the compressed binary \texttt{PDFxTMD-DPDFB1} format. The file includes only momentum-fraction pairs satisfying $x_1+x_2\leq1$. The flavor-exchange and sea-quark symmetries of the GS09-based reference distributions allow a reduced representation comprising 36 regular channels, supplemented by separately stored $s\bar{s}$, $c\bar{c}$, and $b\bar{b}$ distributions. These additional channels preserve same-flavor quark--antiquark correlations arising from the GS09 input construction and subsequent perturbative evolution, which cannot be recovered by simply identifying sea quarks with their antiquarks~\cite{Gaunt:2009re}. For each momentum-fraction pair and flavor channel, the table contains all $N_\mu^2=41^2=1681$ combinations of the two factorization-scale nodes. The grid axes are stored in double precision, while the nodal values are converted to single precision and compressed using Zstandard at compression level 3.

The resulting binary grid occupies approximately $740\;\mathrm{MB}$, compared with about $2.6\;\mathrm{GB}$ for the ASCII representation, corresponding to a reduction in file size by a factor of approximately $3.5$. Although this substantially reduces disk storage, the compressed file size does not represent the runtime memory requirement because the nodal payload must be decompressed before evaluation. These grids therefore serve as the reference data for the hybrid neural-network surrogate introduced in the following sections.

\section{Hybrid Neural-Network Representation of DPDs}
\label{sec:HybridNN}

The reference DPD grid introduced in the previous section provides accurate distributions over the full four-dimensional kinematic domain, but storing all of its nodal values requires considerable disk space and runtime memory. The surrogate reduces this cost without changing the underlying reference calculation or the interpolation procedure used to evaluate the distributions.

A neural network provides a compact way of describing the dependence of the DPDs on $x_1$, $x_2$, $\mu_1^2$, and $\mu_2^2$. Matching every grid node and flavor channel with the neural network alone is much harder than obtaining good overall agreement. It may require a larger and slower model, while still offering no guarantee that isolated difficult regions will satisfy the prescribed tolerance. We therefore use the network to reproduce most of the reference grid and store the original reference values only at the small number of nodes where the neural prediction does not meet the required accuracy.

The grid itself remains central to this construction, and we train the network on the nodal values of the \texttt{PDFxTMD-DPDFB1} reference table. For an off-grid request, PDFxTMD identifies the neighboring nodes required by its interpolation procedure. Their values are reconstructed by the neural model and, where necessary, replaced by stored reference values before the usual interpolation is applied. The surrogate neural network model therefore replaces the explicit storage of most nodal values while preserving the original grid structure and interpolation procedure used for DPD evaluation.

The following subsections describe the construction of the hybrid representation, including the numerical parameterization of the DPDs, the neural-network architecture and training strategy, the sparse correction procedure used to preserve accuracy, and the runtime implementation within the grid-based evaluation framework.

\subsection{Kinematic Representation}
\label{subsec:Feature}

Before training, the four kinematic variables $(x_1,x_2,\mu_1^2,\mu_2^2)$ are transformed into a representation better suited to the numerical structure of the reference DPDs. Both the momentum fractions and factorization scales span several orders of magnitude, so we work with logarithmic coordinates and normalize them to a common interval. We then supplement these basic coordinates with derived features that make important structures of the DPD phase space explicit to the network, including the longitudinal kinematic boundary, the asymmetry between the two parton legs, and the heavy-flavor thresholds.

For a positive variable $q$ defined on the interval $[q_{\min},q_{\max}]$, we introduce the logarithmic mapping
\begin{equation}
	L(q;q_{\min},q_{\max})
	=
	2\frac{\ln(q/q_{\min})}
	{\ln(q_{\max}/q_{\min})}
	-1.
	\label{eq:log_feature}
\end{equation}
This transformation maps the logarithmic range of $q$ onto $[-1,1]$, preserving resolution over several orders of magnitude while placing the different input coordinates on a comparable numerical scale.

The momentum-fraction and factorization-scale coordinates are represented as
\begin{equation}
	\ell_i=L(x_i;x_{\min},x_{\max}),
	\qquad
	\tau_i=L(\mu_i^2;\mu_{\min}^2,\mu_{\max}^2),
	\qquad i=1,2.
\end{equation}
We use $x_{\min}=10^{-6}$, $x_{\max}=0.999$, $\mu_{\min}^2=1\,\mathrm{GeV}^2$, and $\mu_{\max}^2=10^9\,\mathrm{GeV}^2$ as the normalization limits. The sampled inputs are restricted to the original reference-grid nodes and must also satisfy $x_1+x_2\leq 1$.

Although the individual coordinates $\ell_1$ and $\ell_2$ determine the values of $x_1$ and $x_2$, they do not explicitly encode the combined distance of a point from the kinematic boundary $x_1+x_2=1$. Rather than requiring the network to infer this relation solely from the two separate inputs, we provide two derived quantities,
\begin{equation}
	s=x_1+x_2,
	\qquad
	r=\max(1-s,x_{\min}),
\end{equation}
where $s$ is the total longitudinal momentum fraction carried by the two active partons and $r$ represents the remaining longitudinal momentum.

These quantities enter the network through the two additional features
\begin{equation}
	2s-1,
	\qquad
	L(r;x_{\min},1).
\end{equation}
The feature $2s-1$ is a normalized form of the total momentum fraction $s=x_1+x_2$, mapping its range from $[0,1]$ to $[-1,1]$. The second feature, $L(r;x_{\min},1)$, describes the proximity to the kinematic boundary $x_1+x_2=1$. The reason for using a logarithmic representation of $r=1-x_1-x_2$ is that $r$ becomes very small near this boundary, and the logarithm provides enhanced resolution in this region. The lower limit $r\geq x_{\min}$ is imposed only to avoid evaluating the logarithm at zero.

We also include two features that describe the asymmetry between the two parton legs,
\begin{equation}
	A_x=\frac{x_1-x_2}{x_1+x_2},
	\qquad
	A_\mu=\frac{\tau_1-\tau_2}{2}.
\end{equation}
The variable $A_x$ measures the relative imbalance between the two momentum fractions and is naturally bounded between $-1$ and $1$. It vanishes for $x_1=x_2$ and approaches $\pm1$ when one momentum fraction becomes much larger than the other. Similarly, $A_\mu$ measures the difference between the two normalized factorization scales. It vanishes when $\mu_1^2=\mu_2^2$, while its sign indicates which scale is larger and its magnitude reflects how different the two scales are. Additionally, both features change sign when the two parton legs are exchanged.

These quantities form the eight-component vector
\begin{equation}
	\bm{\phi}_8=
	\left(
	\ell_1,\ell_2,L(r;x_{\min},1),
	\tau_1,\tau_2,A_x,A_\mu,2s-1
	\right).
	\label{eq:features8}
\end{equation}

The final four features describe the positions of the charm and bottom thresholds relative to each factorization scale. These features are useful because the behavior of the heavy-flavor channels changes near their activation thresholds. We first define
\begin{equation}
	z_{iq}
	=
	\log_{10}\left(\frac{\mu_i^2}{m_q^2}\right),
	\qquad i=1,2,\quad q=c,b,
\end{equation}
and restrict its range through
\begin{equation}
	h_{iq}
	=
	\begin{cases}
		-1, & z_{iq}<-1,\\
		z_{iq}, & -1\leq z_{iq}\leq 1,\\
		1, & z_{iq}>1.
	\end{cases}
	\label{eq:threshold_features}
\end{equation}
The feature $h_{iq}$ is zero when $\mu_i^2=m_q^2$, negative below the corresponding threshold, and positive above it. It varies only within one decade of the threshold on either side and remains fixed at $-1$ or $1$ outside this region. We use $m_c=1.40\;\mathrm{GeV}$ and $m_b=4.75\;\mathrm{GeV}$, consistently with the reference setup. Appending $(h_{1c},h_{2c},h_{1b},h_{2b})$ to $\bm{\phi}_8$ gives the 12 network inputs.

\subsection{Single-PDF Baseline and Target Representation}
\label{subsec:Target}

After defining the kinematic features used as inputs to the feed-forward neural network, we next consider the numerical representation of the DPD values themselves. The unweighted distributions can span a large dynamic range, particularly toward small momentum fractions. We therefore train the surrogate using the momentum-weighted quantities $xf_a(x,\mu^2)$ and $x_1x_2D_{a_1a_2}(x_1,x_2;\mu_1^2,\mu_2^2)$, which reduce this variation and provide a better-conditioned regression target. For simplicity, the symbols $f$ and $D$ will continue to denote these numerical values throughout the surrogate construction.

Rather than asking the network to learn the complete DPD directly, we construct a baseline from the corresponding single-parton distributions. Much of the dependence on $x_1$, $x_2$, $\mu_1^2$, and $\mu_2^2$ is already contained in the individual PDFs, allowing the network to focus on the additional two-parton structure that distinguishes the correlated reference DPD from a factorized description.

To incorporate the dominant suppression near the longitudinal kinematic boundary, we multiply the single-PDF product by the flavor-independent phase-space factor

\begin{equation}
	\rho_0(x_1,x_2)
	=
	\left[
	\frac{1-x_1-x_2}
	{(1-x_1)(1-x_2)}
	\right]^2,
	\qquad
	x_1+x_2<1.
	\label{eq:boundary_envelope}
\end{equation}
This form corresponds to the flavor-independent part of the phase-space suppression used in the GS09 construction~\cite{Gaunt:2009re}. It approaches zero as $x_1+x_2\rightarrow1$, while tending to unity when either momentum fraction becomes small. The full GS09 input prescription also introduces additional endpoint modifications for valence-quark contributions. In the present surrogate, we deliberately omit these flavor-dependent refinements and use the same factor $\rho_0(x_1,x_2)$ for all flavor channels. This keeps the baseline simple and general, while leaving the remaining flavor-dependent structure to be learned by the neural network.

For a flavor pair $(a_1,a_2)$, we define the baseline as
\begin{equation}
	B_{a_1a_2}
	=
	f_{a_1}(x_1,\mu_1^2)
	f_{a_2}(x_2,\mu_2^2)
	\rho_0(x_1,x_2).
	\label{eq:spdf_baseline}
\end{equation}
This factorized form captures the dominant single-parton dependence together with the leading suppression near the kinematic boundary. Rather than asking the neural network to reproduce the full DPD magnitude directly, we train it on the deviation of the reference distribution from the factorized baseline. We define the relative correction
\begin{equation}
	R_{a_1a_2}
	=
	\frac{
		D_{a_1a_2}^{\mathrm{ref}}
		-
		B_{a_1a_2}
	}
	{
		B_{a_1a_2}
		+
		\varepsilon_{a_1a_2}
	},
	\label{eq:relative_residual}
\end{equation}
where $\varepsilon_{a_1a_2}$ is a small positive channel-dependent numerical scale introduced to keep the relative correction well behaved when the baseline becomes very small. Because the characteristic magnitudes of the different DPD flavor channels can vary substantially, this scale is determined separately for each channel. Its precise definition, including the treatment of exchange-related channels, is given in Appendix~\ref{app:epsilon}.

Although the relative correction $R_{a_1a_2}$ already removes much of the variation associated with the overall magnitude of the DPD, it can still span a wide numerical range, particularly in regions where the reference distribution differs strongly from the baseline. Using $R_{a_1a_2}$ directly as the regression target would allow a relatively small number of large residuals to dominate the optimization. To reduce this dynamic range without losing sensitivity to small corrections, we apply the inverse hyperbolic sine transformation. The neural-network target is defined as
\begin{equation}
	z_{a_1a_2}
	=
	\operatorname{asinh}
	\left(
	R_{a_1a_2}
	\right).
	\label{eq:asinh_target}
\end{equation}
The transformation is approximately linear for $|R|\ll1$, so small deviations from the baseline are preserved, while its logarithmic growth at large $|R|$ compresses extreme corrections.

During inference, the network predicts $\widehat z_{a_1a_2}$ and the corresponding DPD value is reconstructed analytically as
\begin{equation}
	D_{a_1a_2}^{\mathrm{NN}}
	=
	B_{a_1a_2}
	+
	\left(
	B_{a_1a_2}
	+
	\varepsilon_{a_1a_2}
	\right)
	\sinh\left(
	\widehat z_{a_1a_2}
	\right).
	\label{eq:asinh_reconstruction}
\end{equation}

The physical kinematic limit and heavy-flavor activation conditions are enforced separately from the neural approximation. Values outside the region $x_1+x_2<1$ are set exactly to zero, while charm and bottom channels are set to zero whenever the corresponding factorization scale lies below the relevant heavy-flavor threshold. These analytically inactive values are not used to define the learned correction.

The resulting target combines a simple physically motivated single-PDF baseline with a relative neural correction. The baseline supplies the dominant single-parton dependence and the common suppression near the longitudinal kinematic boundary, while the neural network learns the remaining non-factorized structure required to reproduce the reference DPD.

\subsection{Neural-Network Architecture}
\label{subsec:NNArchitecture}

The transformed targets defined in the previous subsection are represented by a single feed-forward neural network shared by all flavor channels. The network receives the 12-component kinematic feature vector described above and returns one output for each of the 121 parton pairs. Thus, a single forward evaluation produces the complete set of transformed DPD corrections at a given point $(x_1,x_2,\mu_1^2,\mu_2^2)$.

The network is a fully connected multilayer perceptron with three hidden layers of width 256. Denoting the input feature vector by $\bm{\phi}_{12}$, the hidden representation is constructed recursively as
\begin{equation}
	\bm{h}^{(0)}
	=
	\bm{\phi}_{12},
\end{equation}
and
\begin{equation}
	\bm{h}^{(\ell)}
	=
	\operatorname{SiLU}
	\left(
	W^{(\ell)}\bm{h}^{(\ell-1)}
	+
	\bm{b}^{(\ell)}
	\right),
	\qquad
	\ell=1,2,3,
	\label{eq:nn_hidden_layers}
\end{equation}
where each hidden vector $\bm{h}^{(\ell)}$ has dimension 256. The SiLU activation function~\cite{Elfwing2018} is defined as
\begin{equation}
	\operatorname{SiLU}(u)
	=
	\frac{u}{1+e^{-u}}.
\end{equation}
During training, dropout~\cite{Srivastava2014} with probability $p=0.01$ is applied after each hidden-layer activation. Dropout is disabled during validation and inference.

The output layer is linear,
\begin{equation}
	\widehat{\bm{z}}
	=
	W^{(\mathrm{out})}
	\bm{h}^{(3)}
	+
	\bm{b}^{(\mathrm{out})},
	\qquad
	\widehat{\bm{z}}\in\mathbb{R}^{121},
	\label{eq:nn_output}
\end{equation}
where each component $\widehat z_{a_1a_2}$ is the relative-asinh target of one flavor pair defined in Eq.~\eqref{eq:asinh_target}. No output activation is applied, since the transformed correction can take either positive or negative values and is not restricted to a finite interval. The complete architecture is $12 \longrightarrow 256 \longrightarrow 256 \longrightarrow 256 \longrightarrow 121$.

The architecture is deliberately kept compact because inference speed is important for the hybrid evaluator. A general off-grid request uses two adjacent nodes along each of the four grid axes and may require neural predictions at up to $2^4=16$ neighboring nodes. The native implementation can process these node inputs as a batch, but a larger or more complex network would still increase the cost of each DPD evaluation. The three-layer, 256-wide MLP provides a compromise between approximation capacity and fast inference. Additionally, the physics-informed formulation also makes this compact architecture possible. The baseline incorporates the dominant endpoint behavior, and the kinematic limit and heavy-flavor activation conditions are imposed analytically. The network is left to learn only the remaining multidimensional correction to the reference DPDs.

\subsection{Training and Sampling Strategy}
\label{subsec:Training}

The neural network is implemented and trained with PyTorch~\cite{Paszke:2019PyTorch} using the tabulated reference DPD grid. Rather than generating continuous random phase-space points, we sample nodes from the same $(x_1,x_2,\mu_1^2,\mu_2^2)$ grid used by the reference representation. This choice makes the training procedure consistent with the subsequent hybrid construction, where the neural network is evaluated at the original grid nodes and inaccurate values are replaced by sparse reference entries.

For the results presented in this work, $10^6$ grid nodes are selected for training and $2\times10^5$ independent nodes for validation. Since DPDs satisfy an exchange symmetry under interchange of the two parton legs, $(x_1,x_2,\mu_1^2,\mu_2^2)\leftrightarrow(x_2,x_1,\mu_2^2,\mu_1^2)$, exchange-related configurations do not provide independent information. Therefore, they are always assigned to the same split when constructing the training and validation sets. If one ordering of a configuration belongs to the training set, its exchanged configuration is also kept in the training set, and similarly for validation. This prevents the network from being trained on one ordering of a configuration while being validated on its symmetry-related counterpart.

After the disjoint training and validation sets have been constructed, data augmentation based on the exchange symmetry is applied separately within each set. For every sampled point, the exchanged configuration is added together with the corresponding reversal of the ordered flavor indices, $z_{a_1a_2}(x_1,x_2,\mu_1^2,\mu_2^2) \longrightarrow z_{a_2a_1}(x_2,x_1,\mu_2^2,\mu_1^2)$.
This procedure doubles the effective number of examples used for training and validation to $2\times10^6$ and $4\times10^5$, respectively.

The reference DPD values and the corresponding single-PDF baseline values are evaluated at each selected node, from which the relative-asinh targets of Eq.~\eqref{eq:asinh_target} are constructed. 
Also, in order to improve accuracy in difficult regions of phase space, additional emphasis is introduced through the training loss. In particular, configurations close to the longitudinal kinematic endpoint receive the smooth sample weight
\begin{equation}
	w_{\mathrm{edge}}(s)
	=
	1+
	\left(w_{\max}-1\right)
	\sigma\left[
	\frac{4(s-s_0)}{\Delta_s}
	\right],
	\qquad
	s=x_1+x_2,
	\label{eq:edge_sample_weight}
\end{equation}
where $\sigma(t)=1/(1+e^{-t})$. We use $s_0=0.65$, $\Delta_s=0.05$, and $w_{\max}=10$.
Thus, points well inside the physical domain have approximately unit weight, whereas configurations approaching the endpoint receive progressively larger importance.

Because the different flavor channels are not equally difficult to approximate, additional channel-dependent weights are introduced in the mean-squared-error loss for the transformed targets $z_{a_1a_2}$. Channels containing at least one gluon are assigned a weight of $3.5$, the gluon--gluon channel a weight of $6$, channels containing charm a weight of $1.5$, and channels containing bottom a weight of $5$. These values were chosen empirically based on the observed training difficulty of the corresponding channel groups, with larger weights assigned to channels that required greater emphasis during optimization. If a channel belongs to more than one category, the largest applicable weight is used.

Heavy-flavor threshold regions are also emphasized locally. For a heavy flavor $h=c,b$ on leg $i$, we define the logarithmic distance from the activation threshold as
\begin{equation}
	d_{ih}
	=
	\left|
	\log_{10}
	\left(
	\frac{\mu_i^2}{m_h^2}
	\right)
	\right|.
\end{equation}
A proximity factor is then defined by
\begin{equation}
	p_{ih}
	=
	\exp\left[
	-\left(
	\frac{d_{ih}}{\Delta_h}
	\right)^2
	\right],
	\qquad
	\Delta_h=0.2.
\end{equation}
This quantity reaches unity at the corresponding heavy-flavor threshold and decreases smoothly away from it. The threshold contribution to the training weight is constructed as
\begin{equation}
	w_{ih}^{\mathrm{thr}}
	=
	1+
	\left(
	w_{\max}^{\mathrm{thr}}-1
	\right)
	p_{ih},
	\qquad
	w_{\max}^{\mathrm{thr}}=4.
\end{equation}
For a sampled point $n$ and flavor channel $k$, $w_{nk}^{\mathrm{thr}}$ is obtained by taking the largest applicable threshold weight over the heavy flavors and the two legs.
The weight varies smoothly from unity away from the threshold to a maximum value of $4$ at $\mu_i^2=m_h^2$. This gives additional training emphasis to the regions where heavy-flavor channels become active.

The training objective is a weighted mean-squared error between the predicted and reference transformed targets. Let $n$ label the sampled phase-space points and $k$ the flavor channels. For each pair $(n,k)$, we define the combined weight $W_{nk} = w_{\mathrm{edge}}(s_n)\, w_{k}^{\mathrm{flav}}\, w_{nk}^{\mathrm{thr}}$. Here, $w_{\mathrm{edge}}(s_n)$ is the endpoint weight, $w_{k}^{\mathrm{flav}}$ is the channel-dependent flavor weight, and $w_{nk}^{\mathrm{thr}}$ is the heavy-flavor threshold weight.

The corresponding loss for the transformed targets $z_{nk}$ is
\begin{equation}
	\mathcal{L}_{z}
	=
	\frac{
		\displaystyle\sum_{n,k}
		W_{nk}
		\left(
		\widehat{z}_{nk}-z_{nk}
		\right)^2
	}{
		\displaystyle\sum_{n,k}
		W_{nk}
	}.
	\label{eq:target_space_loss}
\end{equation}
Heavy-flavor entries that are inactive at the corresponding factorization scales are excluded from the loss, since their vanishing values are imposed analytically rather than learned by the neural network.

Training is performed with AdamW~\cite{Loshchilov:2019AdamW} using minibatches of 4096 samples, an initial learning rate of $4\times10^{-3}$, and weight decay $10^{-6}$. The model is optimized for 600 epochs. A cosine-annealing schedule~\cite{Loshchilov:2017SGDR} gradually reduces the learning rate to $5\%$ of its initial value by the end of training. The Euclidean norm of the gradient is limited to a maximum value of $5$ before each optimizer update to suppress occasional large updates from difficult phase-space regions.

Model selection is based on the weighted mean-squared-error loss of Eq.~\eqref{eq:target_space_loss}, and the checkpoint with the lowest finite validation loss is selected for deployment.

\subsection{Hybrid Grid Construction and Sparse Corrections}
\label{subsec:HybridGrid}

After training, the neural network is combined with a sparse set of reference DPD values to construct the hybrid grid. The hybrid uses the same $N_x=56$ momentum-fraction nodes and $N_\mu=41$ factorization-scale nodes as the reference table. We then scan every momentum-fraction pair in the physical region of the reference grid, all combinations of the two scale nodes, and each of the 121 ordered flavor channels. At every such combination, the network prediction is converted back to the DPD representation using Eq.~\eqref{eq:asinh_reconstruction} and compared with the value read directly from the reference table. The comparison uses the DPD value stored at each grid point, rather than one reconstructed by applying the reference-grid interpolation at that same point.

The acceptance criterion uses the larger of relative and absolute tolerances. The relative tolerance controls the error for DPD values of ordinary size, while the absolute tolerance prevents very small reference values from producing an excessively restrictive relative error. To account for the different characteristic magnitudes of the flavor channels, we define a separate certification scale. For channel $k$, its initial value is $\Lambda_k^{(0)}=Q_{0.90}\left(\left|D_{nk}^{\mathrm{ref}}\right|\right)$, where the percentile is evaluated over the training sample after the longitudinal kinematic limit and heavy-flavor conditions have been imposed. Exchange-related channels are assigned the same certification scale, $\Lambda_{a_1a_2}=\Lambda_{a_2a_1}=\max\left(\Lambda_{a_1a_2}^{(0)},\Lambda_{a_2a_1}^{(0)}\right)$. This scale is used only for the construction and certification of the hybrid grid. It is distinct from the stabilization scale $\varepsilon_{a_1a_2}$ entering the relative-asinh target.

For each physical four-dimensional grid node $n$ and active ordered flavor channel $k$, we define the normalized nodal error as
\begin{equation}
	\mathcal{E}_{nk}
	=
	\frac{
		\left|
		D_{nk}^{\mathrm{NN}}
		-
		D_{nk}^{\mathrm{ref}}
		\right|
	}{
		\max\left(
		\alpha\Lambda_k,
		r\left|D_{nk}^{\mathrm{ref}}\right|
		\right)
	},
	\label{eq:hybrid_nodal_error}
\end{equation}
with $\alpha=10^{-5}$ and $r=0.02$.
The neural prediction is accepted when $\mathcal{E}_{nk}\leq1$. Equivalently, the permitted absolute deviation is
\begin{equation}
	\left|
	D_{nk}^{\mathrm{NN}}
	-
	D_{nk}^{\mathrm{ref}}
	\right|
	\leq
	\max\left(
	10^{-5}\Lambda_k,
	0.02\left|D_{nk}^{\mathrm{ref}}\right|
	\right).
	\label{eq:hybrid_acceptance}
\end{equation}
Thus, the criterion gives a $2\%$ relative tolerance when the reference value is large compared with the channel scale, while the first branch provides a channel-dependent absolute tolerance when the reference value is close to zero.

Whenever Eq.~\eqref{eq:hybrid_acceptance} is not satisfied, the reference nodal value is written to the sparse correction table. The hybrid nodal value is then
\begin{equation}
	D_{nk}^{\mathrm{hyb}}
	=
	\begin{cases}
		D_{nk}^{\mathrm{NN}},
		& \mathcal{E}_{nk}\leq1,\\
		D_{nk}^{\mathrm{ref}},
		& \mathcal{E}_{nk}>1.
	\end{cases}
	\label{eq:hybrid_nodal_value}
\end{equation}
Each sparse entry records the four grid indices, the ordered flavor-channel index, and the reference value. These entries are replacements, and when an entry is present, the reference value replaces the neural prediction before interpolation.

For a general phase-space point, PDFxTMD first identifies the neighboring grid nodes required by its interpolation procedure. Their values are supplied by the neural network and replaced by sparse reference entries where necessary, after which the standard interpolation is applied. The network reconstructs only the nodal values and is not evaluated directly at the requested off-grid coordinate.

\subsection{Native Representation and Runtime Evaluation}
\label{subsec:NativeRuntime}

The trained network and sparse correction table are converted into a native hybrid representation implemented within PDFxTMDLib~\cite{Valeshabadi2026}. The hybrid representation uses the \texttt{PDFxTMD-DPDFH1} set format, while the dense reference grid remains available in the \texttt{PDFxTMD-DPDFB1} format. Both representations are selected through the PDFxTMD factory and accessed through the same public DPD interface, with the argument order $(a_1,a_2,x_1,\mu_1^2,x_2,\mu_2^2)$. They both return the momentum-weighted quantity $x_1x_2D_{a_1a_2}$.

The neural network surrogate model is developed and trained in Python using PyTorch~\cite{Paszke:2019PyTorch}. After training, the checkpoint is exported to a framework-independent file containing the kinematic domain, heavy-flavor masses, stabilization scales, flavor ordering, and network operations and parameters. The network parameters are stored in single precision and evaluated directly in C++ with oneDNN~\cite{oneDNN}. As a result, the evaluation stage requires neither Python nor PyTorch. Together with the neural model, the deployed hybrid representation contains the grid information, the identifier of the single-PDF set used in the baseline construction, and the sparse reference corrections required to reconstruct the hybrid nodal values according to Eq.~\eqref{eq:hybrid_nodal_value}.

When the hybrid set is initialized, the specified single-PDF set is loaded and its values on the $(x,\mu^2)$ grid are kept in memory. For an arbitrary request, PDFxTMD first identifies the neighboring nodes required by its interpolation procedure. At each required node, the single-PDF values and the common phase-space factor of Eq.~\eqref{eq:boundary_envelope} are used to construct the baseline of Eq.~\eqref{eq:spdf_baseline}. The native network then supplies all 121 transformed outputs, which are converted to DPD values using Eq.~\eqref{eq:asinh_reconstruction}. The longitudinal kinematic constraint and heavy-flavor conditions are imposed explicitly, and any sparse entry replaces the neural value before interpolation. The resulting nodal values are passed to the same interpolation prescription used for the dense reference grid, including its treatment of the $x_1+x_2=1$ boundary and interpolation in $\log\mu^2$.

To avoid repeated neural inference, complete 121-channel nodal results may be placed in a least-recently-used cache. The result for the most recently requested kinematic point is also cached, so that several flavor combinations at the same coordinates can reuse one neural reconstruction and interpolation. The native evaluator supports multithreaded execution when reconstructing the nodal values required for an interpolation. The preferred number of execution threads and the maximum cache capacity are stored in the metadata of the DPD set. These settings control the computational performance and memory use of the runtime.

\section{Numerical Validation and Performance}
\label{sec:NumVal_Perf}

This section evaluates the hybrid representation in terms of numerical accuracy, resource requirements, and computational performance. All benchmarks are performed on a system equipped with an 11th-generation Intel Core i9-11900K processor, with 8 physical cores and 16 hardware threads, and $32\;\mathrm{GiB}$ of memory, running Linux Mint 22.3. Unless stated otherwise, the hybrid evaluator uses four execution threads. Although the final set metadata specifies a node-cache capacity of $50\,000$ for ordinary use, the persistent node cache is disabled for all performance and memory measurements. This no-cache setup prevents reuse of nodal values reconstructed for earlier phase-space points and gives a conservative performance test. Each test uses the final packaged hybrid set rather than the trained model alone. The hybrid and dense reference sets are loaded through the same PDFxTMD interface, so the comparison covers the complete evaluation chain: construction of the single-PDF baseline, native neural inference, sparse nodal replacements, and interpolation.

\subsection{Accuracy Validation}
\label{subsec:Evaluation}

The exhaustive scan described in Sec.~\ref{subsec:HybridGrid} sets the accuracy of the hybrid representation at the original grid nodes. As it is already discussed, every active nodal value that does not satisfy Eq.~\eqref{eq:hybrid_acceptance} is replaced by its reference value. For the model used here, the scan covers $4\,879\,943$ physical grid points and $554\,383\,007$ active ordered flavor values. Only $3\,845\,458$ reference values, or $0.694\%$ of the active combinations, are needed as replacements. The neural network supplies the remaining values within the prescribed mixed absolute--relative tolerance.

\begin{figure*}[htbp]
	\centering
	\includegraphics[width=0.96\textwidth]{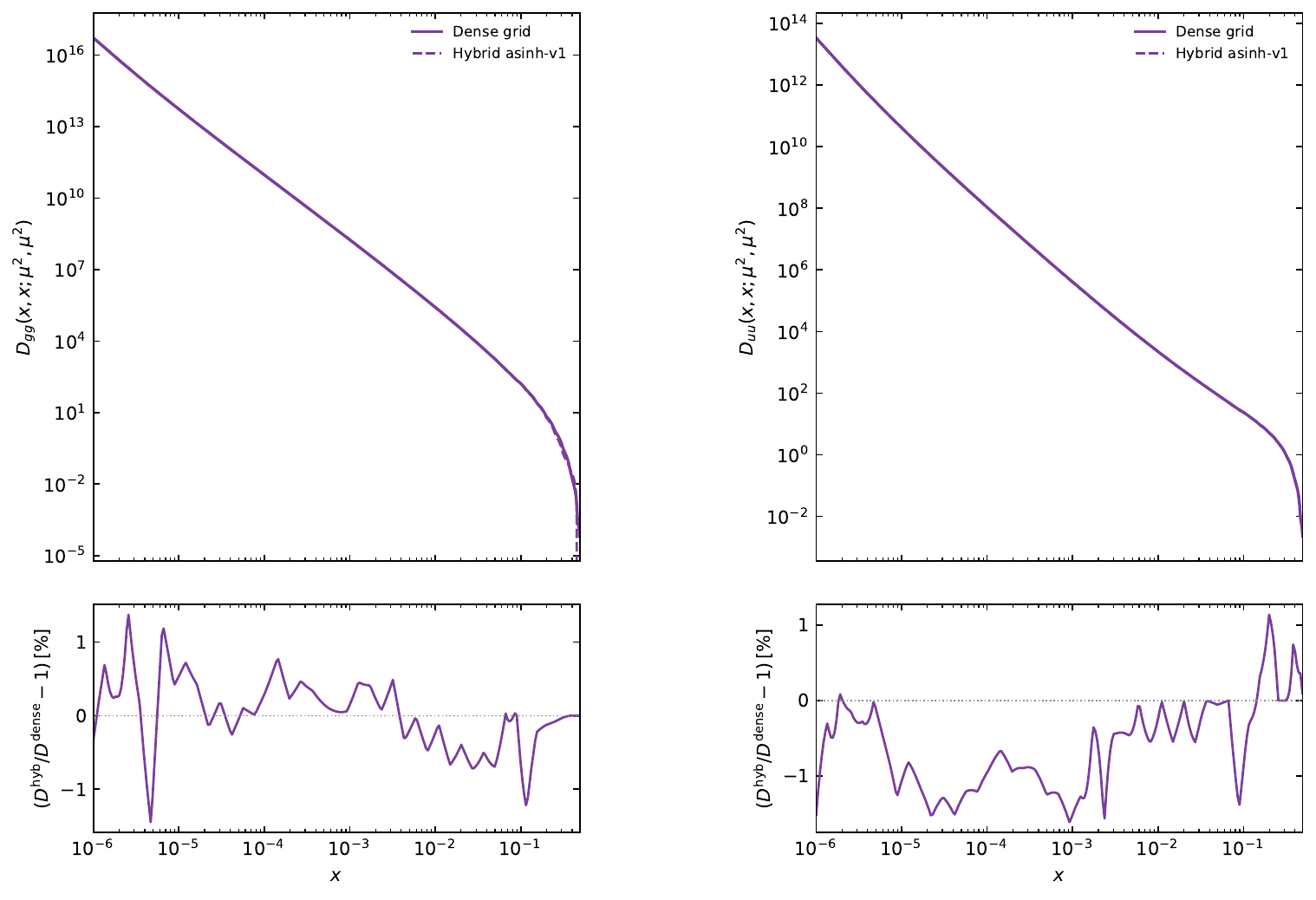}
	\caption{Comparison of the dense-grid and hybrid representations for $D_{gg}(x,x;\mu^2,\mu^2)$ and $D_{uu}(x,x;\mu^2,\mu^2)$ at $\mu^2=10\;\mathrm{GeV}^2$. The lower panels show the percentage difference of the hybrid result relative to the dense-grid result.}
	\label{fig:dpdf_comparison}
\end{figure*}

As an example of the agreement after the complete runtime reconstruction and interpolation procedure, Fig.~\ref{fig:dpdf_comparison} compares the dense and hybrid results along the diagonal $x_1=x_2=x$ and $\mu_1^2=\mu_2^2=10\;\mathrm{GeV}^2$. We show the gluon--gluon and up--up channels. The upper panels display the DPDs themselves, while the lower panels show the percentage difference between the hybrid and dense results. The two representations follow the same behavior over the full plotted range, including the strong suppression near the longitudinal boundary.

\begin{figure*}[htbp]
	\centering
	\includegraphics[width=0.98\textwidth]{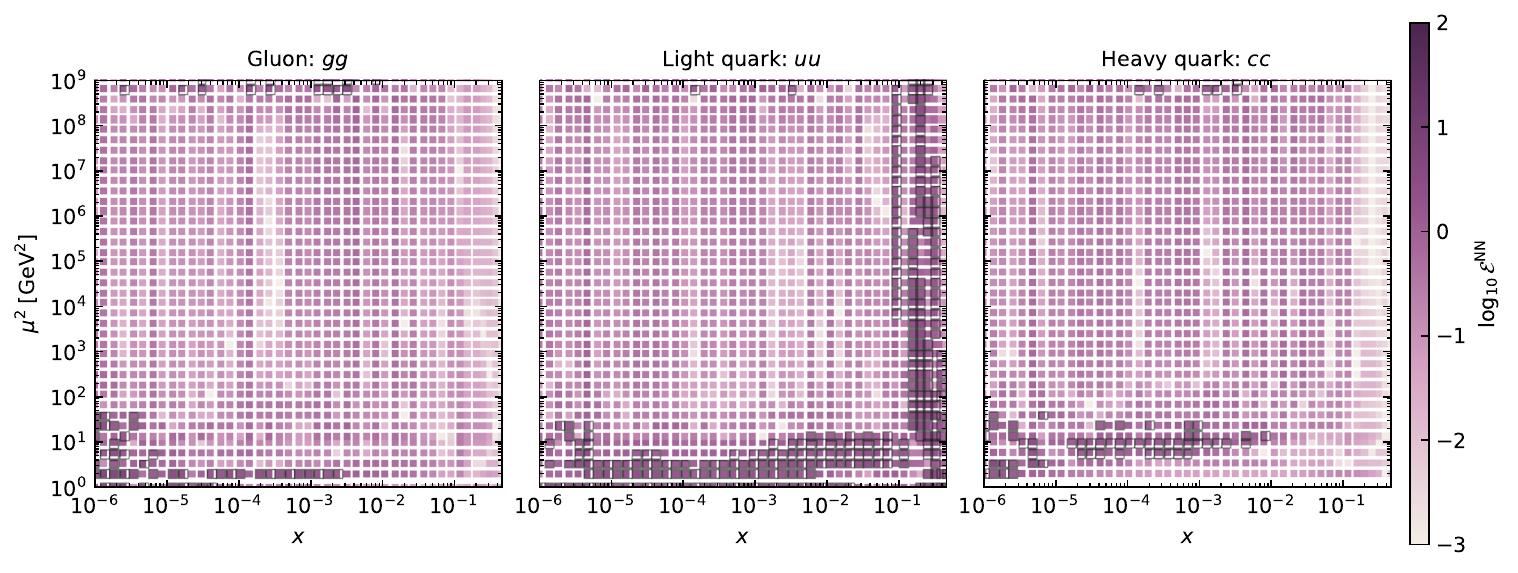}
	\caption{Neural-network error measure in the $(x,\mu^2)$ plane at the diagonal nodes of the reference grid, with $x_1=x_2=x$ and $\mu_1^2=\mu_2^2=\mu^2$, for the $gg$, $uu$, and $cc$ channels. The comparison uses the stored reference nodal values before sparse replacement and does not involve interpolation. Dark outlines identify nodes with $\mathcal{E}_{nk}>1$, which are supplied by the sparse correction table. Values at or below the lower color limit correspond to $\mathcal{E}_{nk}\leq10^{-3}$; the inactive charm region below threshold is omitted.}
	\label{fig:nodal_error_heatmap}
\end{figure*}

To examine the neural component independently of interpolation and sparse replacement, Fig.~\ref{fig:nodal_error_heatmap} compares its predictions directly with the stored reference values on the diagonal grid defined by $x_1=x_2=x$ and $\mu_1^2=\mu_2^2=\mu^2$. The horizontal and vertical axes contain only the original momentum-fraction and scale nodes, respectively. Representative gluon, light-quark, and heavy-quark channels are selected to illustrate different structures of the reference DPDs. The color represents the normalized error $\mathcal{E}_{nk}$ of Eq.~\eqref{eq:hybrid_acceptance}, while dark outlines mark nodes with $\mathcal{E}_{nk}>1$ that are supplied by the sparse correction table. Their distribution shows where the network alone does not meet the prescribed tolerance.

The larger concentration of sparse replacements in the $uu$ channel is consistent with the non-factorized valence-number structure of the GS09 input distributions. In particular, the equal-flavor up-valence contribution contains a number-effect subtraction reflecting the finite number of up-valence quarks in the proton, in addition to the flavor-dependent phase-space structure of the input DPDs~\cite{Gaunt:2009re}. These features are not explicitly encoded in the common single-PDF baseline and therefore increase the complexity of the function represented by the neural network. The observed channel dependence may therefore arise from both differences in the underlying physical structure of the DPD channels and variations in the difficulty of approximating them with the neural model.

\subsection{Resource Requirements and Computational Performance}
\label{subsec:StorageMemory_Perfomance}

\begin{figure}[htbp]
	\centering
	\includegraphics[width=\columnwidth]{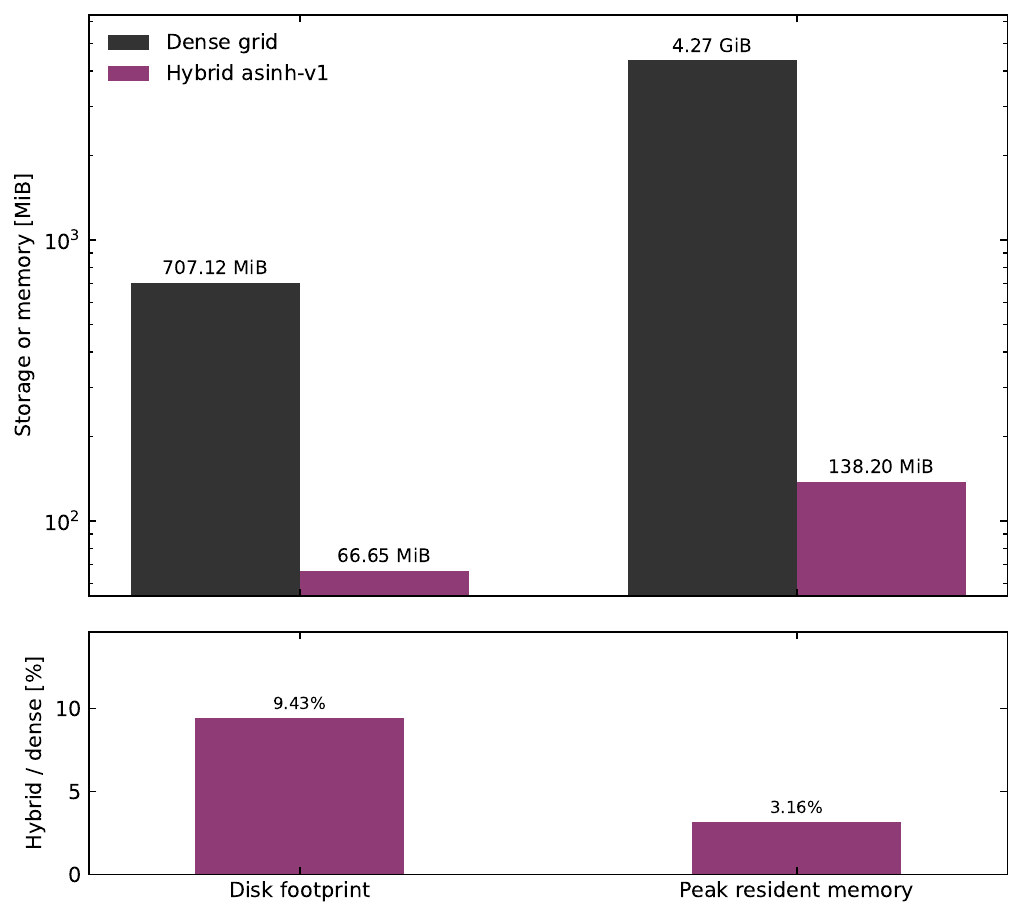}
	\caption{Disk footprint and median peak resident memory of the dense-grid and hybrid representations in the no-cache benchmark. The lower panel gives the hybrid resource requirement as a percentage of the dense result.}
	\label{fig:resource_footprint}
\end{figure}

Beyond numerical accuracy, the main purpose of the hybrid construction is to reduce the storage and memory cost of multidimensional DPD grids. As shown in Fig.~\ref{fig:resource_footprint}, the dense-grid member occupies $707.12\;\mathrm{MiB}$ on disk after its single-precision nodal values are compressed with Zstandard, as described in Sec.~\ref{sec:DPD_Models}. The complete hybrid member occupies $66.65\;\mathrm{MiB}$, giving a reduction by a factor of $10.6$ relative to this compressed dense grid. 

In the no-cache benchmark, the median peak resident memory is $138.20\;\mathrm{MiB}$ for the hybrid and $4.27\;\mathrm{GiB}$ for the dense representation, a reduction by a factor of $31.6$.

The computational performance is measured through the native C++ PDFxTMD interface. The benchmark generates reproducible random physical points with logarithmically sampled momentum fractions and scales and requests all 121 ordered flavor combinations at each point. Initialization time is measured separately from the evaluation loop. Since a single neural evaluation naturally produces all flavor channels, the complete 121-channel result for the current kinematic point remains available while those channels are requested.

The dense and hybrid measurements use the same kinematic points, flavor requests, and execution environment. Several workload sizes are tested to verify that the measured throughput is stable for sufficiently large evaluation samples. Each measurement is repeated independently seven times, and the median and interquartile range are used to summarize the results.

\begin{figure}[htbp]
	\centering
	\includegraphics[width=\columnwidth]{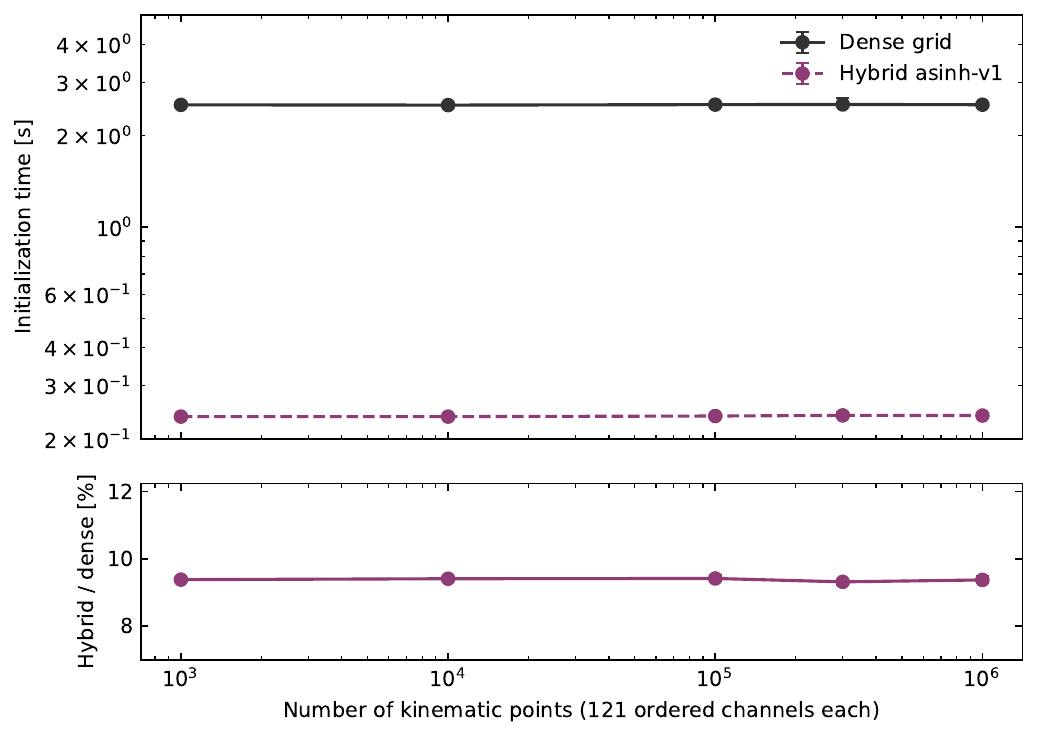}
	\caption{Initialization time of the dense-grid and hybrid representations as a function of the subsequent evaluation workload. Points show medians over seven independent runs and error bars show the interquartile range. The lower panel gives the hybrid time as a percentage of the dense result.}
	\label{fig:initialization_time}
\end{figure}

Figure~\ref{fig:initialization_time} shows that the initialization time is reduced from approximately $2.53\;\mathrm{s}$ for the dense grid to $0.238\;\mathrm{s}$ for the hybrid representation, a reduction by a factor of about $10.6$. This difference arises mainly because the dense representation must decompress the full nodal payload into memory, whereas the hybrid representation loads the substantially smaller native network and sparse correction data.

\begin{figure}[htbp]
	\centering
	\includegraphics[width=\columnwidth]{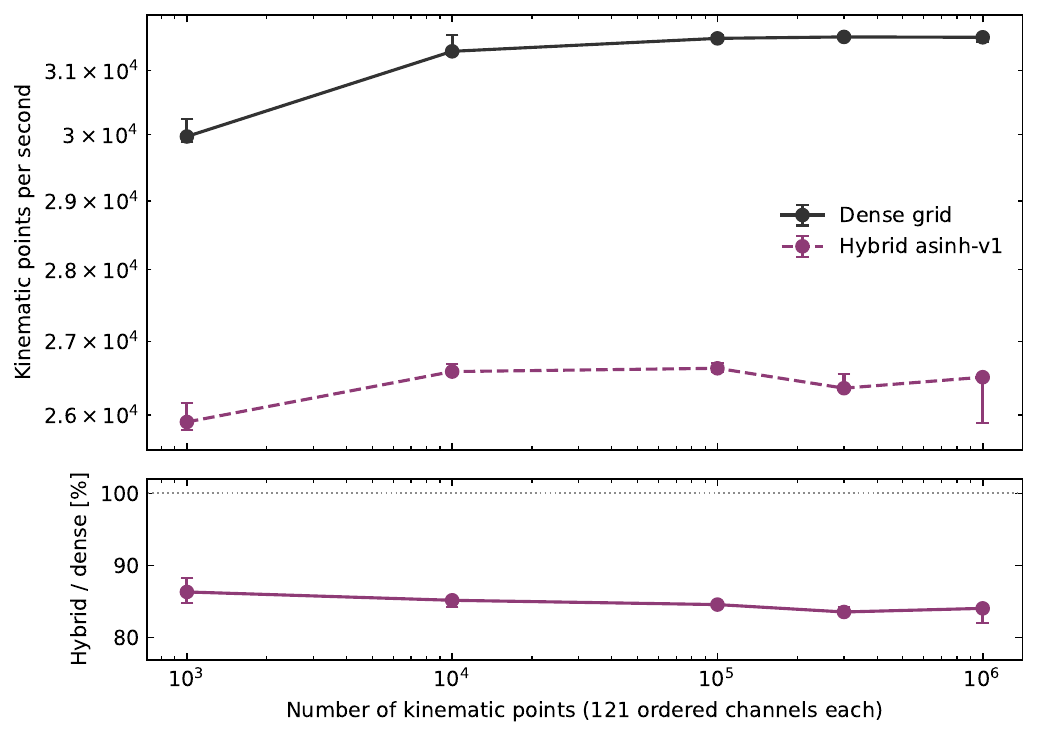}
	\caption{Throughput for evaluations of all 121 ordered flavor channels at each kinematic point. Points show medians over seven independent runs and error bars show the interquartile range. The lower panel gives the hybrid throughput as a percentage of the dense result.}
	\label{fig:evaluation_throughput}
\end{figure}

The evaluation throughput is shown in Fig.~\ref{fig:evaluation_throughput}. The dense interpolator evaluates approximately $3.1\times10^4$ full-flavor kinematic points per second, while the hybrid evaluates approximately $2.6\times10^4$. Its throughput is about $85\%$ of the dense-grid result for this workload. These results show that the main computational advantage of the hybrid representation lies in its much smaller initialization cost and memory requirement, while maintaining a comparable evaluation rate.

\begin{figure}[htbp]
	\centering
	\includegraphics[width=\columnwidth]{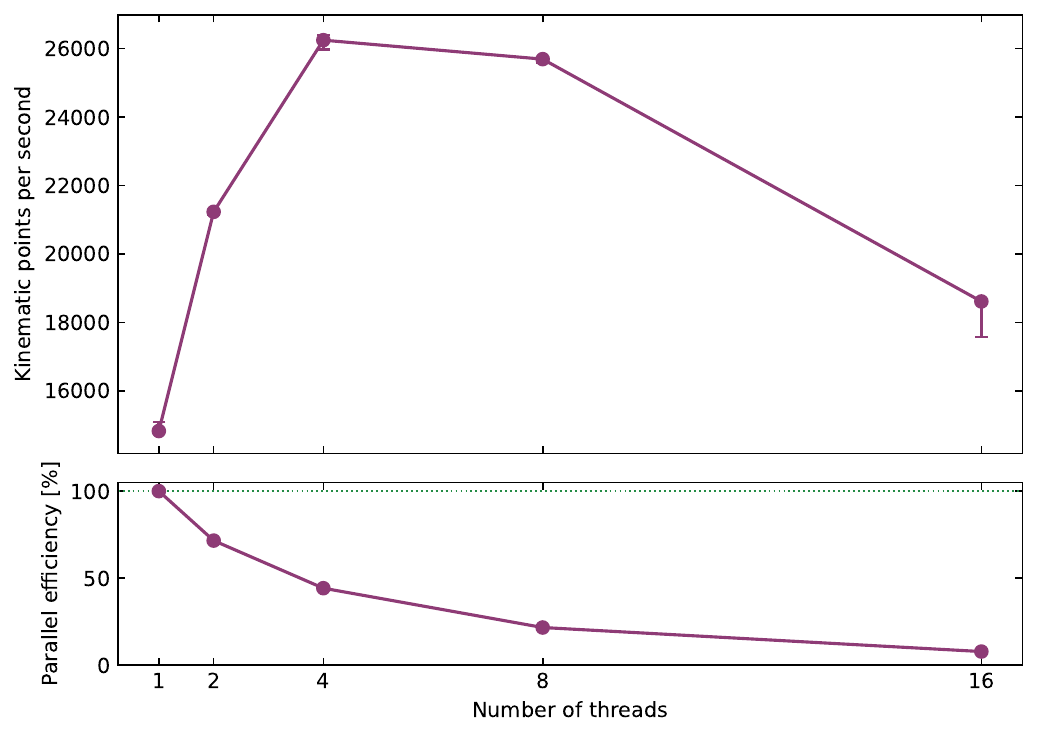}
	\caption{Hybrid evaluation throughput as a function of the number of execution threads for $10^5$ kinematic points, with all 121 ordered flavor channels requested at each point. Points show medians over seven independent runs and error bars show the interquartile range. The lower panel gives the parallel efficiency as a percentage of ideal linear scaling.}
	\label{fig:thread_scaling}
\end{figure}

The effect of the number of execution threads is examined separately for the hybrid model using $10^5$ kinematic points and the same all-channel evaluation procedure. For this test, only the thread-count entry in the set metadata is changed, taking values of 1, 2, 4, 8, and 16, while all other settings remain fixed. Each configuration is measured seven times. As shown in Fig.~\ref{fig:thread_scaling}, the median throughput increases from approximately $1.48\times10^4$ points per second with one thread to $2.62\times10^4$ points per second with four threads, corresponding to a speedup of $1.77$. Increasing the thread count beyond four does not improve the throughput for this workload. The performance is slightly lower with eight threads and decreases to approximately $1.86\times10^4$ points per second with sixteen threads. The maximum throughput is obtained with four threads on the tested hardware. The lower panel of Fig.~\ref{fig:thread_scaling} shows the corresponding parallel efficiency, defined as $\eta(N)=R(N)/(N R(1))$, where $R(N)$ denotes the throughput obtained with $N$ threads. The decrease in efficiency with increasing thread count indicates that additional threads provide diminishing returns for this evaluation workload.

\section{Conclusions}
\label{sec:Conclusions}
In this work, we show that a large four-dimensional DPD table can be compressed by using a feed-forward neural network as a surrogate for its nodal values. Rather than learning the full DPD directly, the network learns the non-factorized correlation structure through relative-asinh corrections to a baseline constructed from single PDFs. The network describes as much of the dense table as possible, while a sparse table stores the original values at nodes where its prediction falls outside the chosen tolerance. During evaluation, the network and sparse corrections reconstruct the nodal values used by the standard interpolation procedure. This avoids loading the complete dense table into memory. Since an off-grid request can involve up to 16 neighboring nodes, the network is deliberately kept compact.

The reference grid has $N_x=56$ nodes along each momentum-fraction axis and $N_\mu=41$ nodes along each scale axis. It contains $4\,879\,943$ physical grid points and $554\,383\,007$ active ordered flavor values. With the $12 \longrightarrow 256 \longrightarrow 256 \longrightarrow 256 \longrightarrow 121$ network used in this study, the exhaustive scan shows that only $3\,845\,458$ of these values, or $0.694\%$, need to be stored as sparse replacements.

The complete hybrid member occupies $66.65\;\mathrm{MiB}$, compared with $707.12\;\mathrm{MiB}$ for the Zstandard-compressed dense-grid member, a reduction by a factor of $10.6$. On the test system, the hybrid also reduces the median peak resident memory from $4.27\;\mathrm{GiB}$ to $138.20\;\mathrm{MiB}$ and the initialization time from about $2.53\;\mathrm{s}$ to $0.238\;\mathrm{s}$. Its throughput is about $85\%$ of the dense-grid rate.

The main gains are in storage, startup time, and memory use, while the evaluation rate remains comparable. The native C++ implementation is available through the same PDFxTMD DPD interface as the dense grid and does not require Python or PyTorch at runtime. Although the present application is limited to $y$-independent DPDs, the same hybrid construction could be extended to distributions with explicit transverse-separation dependence.

\section*{Code and Data Availability}
The source code and DPD set files associated with this work can be downloaded from the PDFxTMDLib DPD webpage at \url{https://pdfxtmdlib.org/dpds/}.

\appendix

\section{Channel-Dependent Stabilization Scale}
\label{app:epsilon}

The relative target defined in Eq.~\eqref{eq:relative_residual} contains the small positive numerical scale $\varepsilon_{a_1a_2}$. Its purpose is to stabilize the normalization in regions where the single-PDF baseline becomes very small.

The characteristic magnitudes of the different DPD flavor channels can differ substantially. A single common value of $\varepsilon$ would regularize large and small channels differently. We instead determine a characteristic scale separately for each ordered flavor pair,
\begin{equation}
	q_{a_1a_2}^{(0)}
	=
	Q_{0.90}
	\left(
	\left|
	D_{a_1a_2}^{\mathrm{ref}}
	\right|
	\right),
	\label{eq:residual_scale_quantile}
\end{equation}
where $Q_{0.90}$ denotes the $90$th percentile evaluated over the physically active training values of that channel.

To preserve the DPD exchange relation, $D_{a_1a_2}(x_1,x_2,\mu_1^2,\mu_2^2)=D_{a_2a_1}(x_2,x_1,\mu_2^2,\mu_1^2)$, the two exchange-related channels are assigned a common numerical scale, $q_{a_1a_2}=q_{a_2a_1}=\max\left(q_{a_1a_2}^{(0)},q_{a_2a_1}^{(0)}\right)$.

The stabilization scale is then defined as $\varepsilon_{a_1a_2}=10^{-4}q_{a_1a_2}$. Because $q_{a_1a_2}$ reflects the typical size of the corresponding flavor channel, this choice scales $\varepsilon_{a_1a_2}$ accordingly and avoids using the same fixed regularization scale for channels with very different magnitudes. The parameter $\varepsilon_{a_1a_2}$ is introduced solely for numerical stability and has no physical interpretation.

\bibliographystyle{apsrev4-2}
\bibliography{references}
\end{document}